\documentclass[11pt,a4paper]{article}
\usepackage{amsfonts}
\usepackage{amsmath}
\usepackage{amssymb}
\usepackage{jheppub}
\usepackage{ulem}
\usepackage{extarrows}
\usepackage{multirow}
\usepackage{float}

\usepackage{inputenc}
\usepackage{xspace}
\usepackage{url}

 \def\be{\begin{equation}}
\def\ee{\end{equation}}
\def\bea{\begin{eqnarray}}
\def\eea{\end{eqnarray}}

\def\a{\alpha}
\def\b{\beta}

\def\s{\sigma}

\def\bp{\bar{p}}

\def\sc{{\hat{c}}}

 \def\p{\partial}

 \def\IZ{\relax\ifmmode\mathchoice
 {\hbox{\cmss Z\kern-.4em Z}}{\hbox{\cmss Z\kern-.4em Z}}
 {\lower.9pt\hbox{\cmsss Z\kern-.4em Z}}
 {\lower1.2pt\hbox{\cmsss Z\kern-.4em Z}}\else{\cmss Z\kern-.4em Z}\fi}
 \def\IB{\relax{\rm I\kern-.18em B}}
 \def\IC{{\relax\hbox{$\inbar\kern-.3em{\rm C}$}}}
 \def\Ic{{\relax\hbox{$\inbar\kern-.22em{\rm c}$}}}
 \def\ID{\relax{\rm I\kern-.18em D}}
 \def\IE{\relax{\rm I\kern-.18em E}}
 \def\IF{\relax{\rm I\kern-.18em F}}
 \def\IG{\relax\hbox{$\inbar\kern-.3em{\rm G}$}}
 \def\IGa{\relax\hbox{${\rm I}\kern-.18em\Gamma$}}
 \def\IH{\relax{\rm I\kern-.18em H}}
 \def\II{\relax{\rm I\kern-.18em I}}
 \def\IK{\relax{\rm I\kern-.18em K}}
 \def\IP{\relax{\rm I\kern-.18em P}}

\def\Tr{{\rm Tr}}
 \font\cmss=cmss10 \font\cmsss=cmss10 at 7pt
 \def\IR{\relax{\rm I\kern-.18em R}}

\def\a{\alpha}
\def\b{\beta}

\def\D{\Delta}

\def\e{\epsilon}

\def\m{\mu}
\def\n{\nu}
\def\s{\sigma}

\def\p{\pi}

\renewcommand{\@}[1]{\sqrt{#1}}
\renewcommand{\le}[1]{\label{#1}\end{eqnarray}}

\def\ffract#1#2{\raise .35 em\hbox{$\scriptstyle#1$}\kern-.25em/
\kern-.2em\lower .22 em \hbox{$\scriptstyle#2$}}

\newdimen\tableauside\tableauside=1.0ex
\newdimen\tableaurule\tableaurule=0.4pt
\newdimen\tableaustep
\def\phantomhrule#1{\hbox{\vbox to0pt{\hrule height\tableaurule width#1\vss}}}
\def\phantomvrule#1{\vbox{\hbox to0pt{\vrule width\tableaurule height#1\hss}}}
\def\sqr{\vbox{%
  \phantomhrule\tableaustep
  \hbox{\phantomvrule\tableaustep\kern\tableaustep\phantomvrule\tableaustep}%
  \hbox{\vbox{\phantomhrule\tableauside}\kern-\tableaurule}}}
\def\squares#1{\hbox{\count0=#1\noindent\loop\sqr
  \advance\count0 by-1 \ifnum\count0>0\repeat}}
\def\tableau#1{\vcenter{\offinterlineskip
  \tableaustep=\tableauside\advance\tableaustep by-\tableaurule
  \kern\normallineskip\hbox
    {\kern\normallineskip\vbox
      {\gettableau#1 0 }%
     \kern\normallineskip\kern\tableaurule}%
  \kern\normallineskip\kern\tableaurule}}
\def\gettableau#1 {\ifnum#1=0\let\next=\null\else
  \squares{#1}\let\next=\gettableau\fi\next}

\newcommand{\auth}{Institute of Theoretical Physics, Aristotle University of Thessaloniki, 54124 Thessaloniki, Greece}

\allowdisplaybreaks
\title{3$d$ Fermion - Boson Map with Imaginary Chemical Potential}
\author[a]{Evangelos G.  Filothodoros,}
\author[a]{\,Anastasios C. Petkou,}
\author[a]{and \,Nicholas D. Vlachos}
\affiliation[a]{\auth}

\emailAdd{efilotho@physics.auth.gr}
\emailAdd{petkou@physics.auth.gr}
\emailAdd{vlachos@physics.auth.gr}
\abstract{We study the three-dimensional $U(N)$ Gross-Neveu and CP$^{N-1}$ models in the canonical formalism with fixed $U(1)$ charge. For large-$N$ this is closely related to coupling the models to abelian Chern-Simons in a monopole background. We show that the presence of the imaginary chemical potential for the $U(1)$ charge makes the phase structure of the models remarkably similar. We calculate their respective large-$N$ free energy densities and show that they are mapped into each other in a precise way. Intriguingly, the free energy map involves the Bloch-Wigner function and its generalizations introduced by Zagier. We expect that our results are connected to the recently discussed $3d$ bosonization.}

\begin{document}
\maketitle
\section{Introduction}

Physics in three dimensions is fascinating, physically relevant and largely experimentally testable. It is therefore  very useful to identify possible universal patterns in it, which eventually  might organize better our understanding of the huge web of condensed matter systems. One such recurrent pattern is 3d bosonization \cite{Fradkin:1994tt}, via  statistical transmutation \cite{Wilczek:1981du,Polyakov:1988md}. Remarkably, but perhaps expectedly, 3d bosonization has been very recently \cite{Karch:2016sxi,Murugan:2016zal,Seiberg:2016gmd,Kachru:2016rui} connected to another recurrent theme of 2+1 dimensional physics,  particle-vortex duality e.g. \cite{Peskin:1977kp}. At finite temperature various dualities between fermionic and bosonic matter theories coupled to non-abelian Chern-Simons have also been recently discussed (see  for example \cite{Giombi:2011kc,Aharony:2012ns} and references therein). Those works present a remarkable progress in our understanding of three-dimensional physics, and its possible holographic higher-spin duals. 

However, if the fermion-boson map is a fundamental property of three-dimensional physics one may wonder whether the presence of non-abelian Chern-Simons gauge fields is necessary to probe it. With this idea in mind we revisit  the finite temperature phase structure of two three-dimensional systems, the fermionic $U(N)$ Gross-Neveu model and the bosonic CP$^{N-1}$ model. We study the systems in the canonical formalism. This can be elegantly done introducing an imaginary chemical potential for the $U(1)$ charge. Nevertheless, it seems that we cannot completely forget about Chern-Simons. The canonical partition function of the systems is intimately related to the  partition function of the same systems coupled to an abelian Chern-Simons gauge field  expanded around a monopole background. For that one probably needs to assume both a suitable mean field approximation \cite{Barkeshli:2014ida} and a large-$N$ expansion. Therefore the system's $U(1)$ charge density is associated to the Chern-Simons level. 

We evaluate the canonical partition function of the models for large-$N$  by a saddle-point expansion. This  leads to a set of gap equations, one for the symmetry breaking order parameter and another for  the charge density. We show that the presence of the imaginary chemical potential essentially maps the phase structure of one system into the other. In particular, the fermionic system gains a finite temperature scaling regime phase which did not exist before, while the bosonic system obtains a continuous  symmetry breaking regime which was previously forbidden by the Mermin-Wagner-Coleman theorem.

Of particular interest is our result for the large-$N$ $U(1)$ charge density of both systems. This is given in terms of the remarkable Bloch-Wigner function \cite{Zagier1} and it turns out to be purely imaginary except at two special values of the chemical potential where it vanishes. The latter property gives rise to two different critical and $U(1)$ neutral vacua for each one of the models. The values of the free energy densities at these vacua exhibit a precise fermion-boson duality, in accordance with expectations. 

Finally, we calculate the on-shell large-$N$ free energy densities of the systems and we show that they can be concisely written in terms of a generalized Bloch-Wigner function introduced by Zagier\cite{Zagier2} (see also \cite{Vanhove} for a recent discussion.) This reveals a precise map between them.  We find that the sum of the free energy densities for values of the imaginary chemical potential that differ by $i\pi$ is given {\it exactly} by the Bloch-Wigner function. 

In Section 2 we discuss some general aspects of three-dimensional systems at imaginary chemical potential and their relation to abelian Chern-Simons coupled to matter. In Section 3 we present our calculations for the gap equations, the phase structure and the free energy density of our models. We conclude and raise a number of open questions on Section 4. The Appendices contain technical details for our calculations in the text.

\section{Imaginary chemical potential for $U(1)$  charge and Chern-Simons theories}

\subsection{Canonical formalism and imaginary chemical potential}

Consider a system at finite temperature $T=1/\beta$ with a global $U(1)$ charge operator $\hat{Q}$. Its canonical partition function can be formally calculated as the thermal average over states with fixed $\hat{Q}$ as
\be
\label{canPF}
Z_c(\b,Q)=\Tr\left[\delta(Q-\hat{Q})e^{-\b\hat{H}}\right]\,.
\ee
We can normalize $Z_c$ dividing with the thermal partition function such that in the absence of charged states $Z_c(\b,0)=1$. One generally expects that the eigenvalues $Q$ of the charge operator $\hat{Q}$ are integers, in which case an explicit  representation for (\ref{canPF}) can be written as 
\be
\label{gcPF}
Z_c(\b,Q)=\int_{0}^{2\pi}\!\frac{d\theta}{2\pi}\,e^{i\theta Q}\,\Tr\left[e^{-\b\hat{H}-i\theta\hat{Q}}\right]=\int_{0}^{2\pi}\!\frac{d\theta}{2\pi}\,e^{i\theta Q}\,Z_{gc}(\b,\mu=-i\theta/\b)\,.
\ee
$Z_{gc}(\b,\mu)$ is the grand canonical partition function with imaginary chemical potential $\mu$. The latter function exhibits in general certain periodicity properties wrt $\theta$ that are intimately connected to the physics of the underlying theory. For example \cite{Roberge} in QCD-like systems with $SU(N)$ non-abelian gauge symmetry and $\hat{Q}$ the fermion number operator, one expects that in the confining phase  the spectrum contains only colour singlets. In this case $Q$ is a multiple of $N$ and $Z_{gc}(\beta,\mu)$ will be periodic  with $\theta$-period  $2\pi /N$. If however there is a phase, e.g. at high temperature, where fundamental particles turn up in the spectrum, one  might expect to find instead a $\theta$-period  $2\pi$.  Indeed, although the $\mathbb{Z}_N$ symmetry of the pure $SU(N)$ Yang-Mills action appears to enforce the $2\pi /N$ periodicity, one generically finds a more complicated structure at high temperature, which may be attributed to a deconfining transition \cite{Aarts:2015tyj}. 

Integrals like (\ref{gcPF}) may be evaluated  by a saddle point analysis. The saddle point equation is 
\be
\label{saddlepoint}
iQ-\b\frac{\partial}{\partial\theta}F_{gc}(\b,-i\theta /\b)= 0\,,
\ee
where the grand canonical potential (i.e. free energy) is $\b F_{gc}(\b,-i\theta/\b)=-\ln Z_{gc}(\b,-i\theta/\b)$. However, in most of the physically relevant situations (i.e. charge conjugation, $CP$ invariance etc), the grand canonical partition function is an even function of $\m$, and hence of $\theta$ \cite{Karbstein:2006er}. Therefore, real solutions for $Q$ would require imaginary $\theta$ and one returns to the usual case of a real chemical potential. Nevertheless, an interesting situation can arise if $F_{gc}(\b,-i\theta/\b)$ has one (or more) extrema for some real $\theta_*$. In such a case the canonical partition function of the system in the absence of charged excitations ($Q=0$) is given, to leading order in some approximation scheme such as large-$N$, by the grand canonical partition function of the same system at fixed imaginary chemical potential $\m_*=-i\theta_*/\b$
\be
\label{approx}
Z_c(\b,0)\approx e^{-\b F_{gc}(\b,-i\theta_* /\b)}\,.
\ee
This is not inconsistent with the normalization of $Z_c$, since  fixing the imaginary chemical potential in a system at finite temperature is tantamount to statistical transmutation \cite{Petkou:1998wd,Christiansen:1999uv}. Despite the absence of charged modes the system described by (\ref{approx}) is in general different from the initial one, as its elementary degrees of freedom would obey different statistics. 

\subsection{Chern-Simons coupled to scalars and fermions in a monopole background}
It is well known\footnote{See for example the standard references \cite{Kapusta:2006pm,ZinnJustin:2002ru}.} that when  scalars and fermions are coupled to a gauge field $A_\m$ at finite temperature, the temporal component $A_0$ is formally equivalent to having an imaginary chemical potential. 
For example, consider Dirac fermions in three Euclidean  dimensions\footnote{Our notations follow \cite{ZinnJustin:2002ru} are briefly presented in the Appendix.} coupled to an abelian Chern-Simons field at level $k$. The finite temperature partition function is
\begin{align}
\label{DiracCSPF}
Z_{f}(\b,k)&=\int [{\cal D}A_\m][{\cal D}\bar\psi][{\cal D}\psi]\exp{\left[-S_f(\bar\psi,\psi,A_\m)\right]}\,,\\
\label{Sf}
S_f(\bar\psi,\psi,A_\m)&=-\int_0^{\beta}\!\!\!d\tau\!\!\int \!\!d^2x\left[\bar{\psi}(\slash\!\!\!\partial -i\slash\!\!\!\!A)\psi+i\frac{k}{4\pi}\epsilon_{\m\n\rho}A_\m\partial_\n A_\rho+...\right]\,,
\end{align}
where the dots denote the possible presence of fermionic self interactions. We expand the CS field around a static (i.e. $\tau$-independent) monopole configuration $\bar{A}_\m$ \cite{Fosco:1998cq}
\be
\label{Aexpansion}
A_\m=\bar{A}_\m+\alpha_\m\,,\,\,\,\bar{A}_\m=(0,\bar{A}_1(x),\bar{A}_2(x))\,,\,\,\, \a_\m=(\a_0(\tau),\a_1(\tau,x),\a_2(\tau,x))\,,
\ee
normalized as\footnote{For example, one may consider the theory on $S^1\times S^2$.}
\be
\label{monopole}
\frac{1}{2\pi}\int d^2x \bar{F}_{12}=1\,,\,\,\,\bar{F}_{\m\n}=\partial_\m \bar{A}_\n-\partial_\n \bar{A}_\n\,.
\ee
Hence, (\ref{Sf}) describes the attachment of $k$ units of monopole charge to the fermions. One then finds
\be
\label{Sfexp}
S_{f}(\bar\psi,\psi,A_\m) =-\int_0^\b \!\!\!d\tau\!\!\int \!\!d^2x\left[\bar\psi(\slash\!\!\!\partial-i\gamma_i\bar{A}_i-i\gamma_\m\a_\m)\psi+i\frac{k}{4\pi}\epsilon_{\m\n\rho}\a_\m\partial_\n\a_{\rho}+..\right]-ik\int_0^\b \!\!\!d\tau a_0\,.
\ee
At this point, we can think of performing the path integral over the CS fluctuations projecting to a sector with fixed total monopole charge. Within this sector we assume the existence of a mean field approximation such that the spatial CS fluctuations compensate for the magnetic background  $\langle \a_i\rangle =-\bar{A}_i$ \cite{Barkeshli:2014ida}. Probably, the validity of such an approximation requires also a suitable large-$N$ limit and it would be interesting to clarify it further.\footnote{A.C.P. wishes to thank Ofer Aharony for an illuminating communication on this point.} We then obtain
\begin{align}
Z_{f}(\b,k)&=\int [{\cal D}\a_0][{\cal D}\bar\psi][{\cal D}\psi]\exp{\left[\int_0^\b \!\!\!d\tau \!\!\int \!\!d^2x\left[\bar\psi(\slash\!\!\!\partial-i\gamma_0\a_0)\psi+..\right]+ik\int_0^\b \!\!\!d\tau\a_0\right]}\nonumber \\
\label{DiracCSPFfin}
&=\int ({\cal D}\theta)e^{ik\theta}Z_{gc,f}(\b,-i\theta/\b)\,,
\end{align}
where in the second line we have defined $\theta=\int_0^\b d\tau\a_0(\tau)$, and compared with the standard formulae in \cite{Kapusta:2006pm,ZinnJustin:2002ru}.  

Similarly, the thermal partition function of a complex scalar $\phi$ coupled to abelian CS at level $k$ may be written as
\begin{align}
\label{ScalarCSPF}
Z_{b}(\b,k)&=\int [{\cal D}A_\m][{\cal D}\bar\phi][{\cal D}\phi]\exp{\left[-S_{b}(\bar\phi,\phi,A_\m)\right]}\,,\\
\label{Ssc}
S_{b}(\bar\psi,\phi,A_\m) &=\int_0^\b\!\! \!d\tau \!\!\int \!\!d^2x \left[|(\partial_\m-iA_\m)\phi|^2 -i\frac{k}{4\pi}\epsilon_{\m\n\rho}A_\m\partial_\n A_\rho+..\right]\,,
\end{align}
where again with the dots we have allowed for the presence of a non trivial scalar potential. Expanding as in (\ref{Aexpansion}), (\ref{monopole}) and assuming a similar mean field  and large-$N$ approximation as above we find
\begin{align}
\label{ScalarCSPFfin}
Z_{b}(\b,k)&=\int[{\cal D}\a_0][{\cal D}\bar{\phi}][{\cal D}\phi]\exp{\left[-\int_0^\b \!\!\!d\tau \!\!\int \!\!d^2x\left[|(\partial_\tau-i\a_0)\phi|^2 +|\partial_i\phi|^2 +..\right]+ik\int_0^\b \!\!d\tau a_0\right]} \nonumber \\
&=\int [{\cal D}\theta]e^{ik\theta}Z_{gc,b}(\b,-i\theta/\b)\,,
\end{align}
where we have used the same definition for $\theta$ as above, and compared with the standard formulae giving the grand canonical partition function for charged scalars \cite{Kapusta:2006pm,ZinnJustin:2002ru}. The above discussion shows that the partition function of fermions and charged scalars coupled to abelian CS in a monopole background are intimately related to their respective canonical partition functions at fixed total $U(1)$ charge.

Now we can also understand the role of the imaginary chemical potential for statistical transmutation. Consider for example the fermionic theory (\ref{DiracCSPFfin}). One notices that the presence of the imaginary chemical potential can be cancelled by the following abelian gauge transformation of the fermions
\be
\label{fgaugetransf}
\psi(\tau,x)\mapsto \psi'(\tau,x)=e^{i\int_0^\tau d\tau'\alpha_0(\tau')}\psi(\tau,x)\,,\,\,\,\bar\psi(\tau,x)\mapsto \bar\psi'(\tau,x)=e^{-i\int_0^\tau d\tau'\alpha_0(\tau')}\bar\psi(\tau,x)\,.
\ee
However, at finite temperature the fermions obey anti-periodic boundary conditions on the thermal circle
\be
\label{apBC}
\psi(\beta,\bar{x})=-\psi(0,\bar{x})\,,\,\,\,\bar{\psi}(\beta,\bar{x})=-\bar{\psi}(0,\bar{x})\,.
\ee
We then see that the gauge transformed fields would satisfy
\be
\label{apBC1}
\psi'(\beta,x)=-e^{i\theta}\psi'(0,x)\,,\,\,\,\,\bar\psi'(\b,x)=-e^{-i\theta}\bar\psi'(0,x)\,.
\ee
Hence, the anti-periodic boundary conditions are preserved only if $\theta=2\pi n$, $n\in \mathbb{Z}$. Other values of $\theta$ would "twist" the boundary conditions and change the  statistical properties of the undelying system. A similar argument goes  through as well for  bosonic systems such as (\ref{ScalarCSPFfin}) where the complex scalars satisfy periodic boundary conditions on the thermal circle. The twisting of the thermal boundary conditions is the main underlying mechanics behind the possible statistical transmutation in systems whose grand canonical potential is extremized at non trivial values of the imaginary chemical potential. 

\section{The fermion-boson map at imaginary chemical potential}

We have argued above that the imaginary chemical potential and its corresponding charge density represent the mean field fluctuation of an abelian Chern-Simons gauge potential around a monopole background. Moreover, one might think that calculating the canonical partition function is systems with  $U(1)$ charges is in principle agnostic to the underlying microscopic structure e.g. whether the elementary degrees of freedom are bosonic or fermionic.  From the outset, the only additional piece of information one has in had is the periodicity of the grand canonical partition function as explained in (\ref{gcPF}). In many ways the situation resembles a quantum mechanical system in a periodic potential. Indeed, looking at (\ref{gcPF}) $\theta$ may be though of as the periodic coordinate and then $Q$ is the analog of the lattice momentum. Such systems usually have a band structure which can be studied by restricting the lattice momentum to the first Brillouin zone. One cannot go very far without a particular microscopic model at hand, however there are certain topological properties like the Zak phase \cite{Zak} which  depend only on the band symmetry.  We will not pursue further this line of ideas here but if there is a lesson to be learned is that there should be some universality in generic canonical partition function calculations for bosonic and fermionic $3d$ systems. This will be exploited below by considering two explicit three dimensional models: the $U(N)$ fermionic Gross-Neveu and the bosonic CP$^{N-1}$  model. 

\subsection{The $U(N)$ Gross-Neveu model at imaginary chemical potential}

To calculate the canonical partition function of the $U(N)$ Gross-Neveu model in the presence of imaginary chemical potential $\m=-i\alpha$ we use the Euclidean action \cite{Petkou:1998wd,Christiansen:1999uv}
\begin{equation}
\label{GNaction}
S_{GN} = -\int_0^\b \!\!\!d\tau\int \!\!d^2x \left[\bar{\psi }^{a}(\slash\!\!\!\partial  -i\gamma_0\alpha)\psi ^{a}+\frac{g}{2N}\left (\bar{\psi }^{a}\psi ^{a}\right )^{2} +iQ\alpha)\right]\,,\,\,\,a=1,2,..N\,.
\end{equation}
where $Q$ is the eigenvalue of the fermion number density\footnote{Wrt  the spatial volume $V_2$.} operator $\hat{Q}=\psi^\dagger\psi$. Introducing an auxiliary scalar field $\sigma$ the canonical partition function is given by
\begin{align}
\label{GNPF1}
Z_f(\b,Q)&=\int({\cal D}\a)({\cal D}\sigma)e^{-S_{f,eff}}\,,\\
\label{GNPF2}
S_{f,eff}&=iQ\int_0^\b \!\!\!d\tau\!\!\int\!\!d^2x\,\alpha -\frac{N}{2g}\int_0^\b \!\!\!d\tau\!\!\int d^2x\,\sigma^2+N\Tr\ln\left(\slash\!\!\!\partial-i\gamma_0\alpha+\sigma\right)_\b\,.
\end{align}
To evaluate (\ref{GNPF1}) we look for constant saddle points $\alpha_*$ and $\sigma_*$. At large-$N$ these are given by the gap equations
\begin{align}
\label{GNgap1}
\frac{\sigma_*}{g}&=\frac{2\sigma_*}{\b}\sum_{n=-\infty}^\infty\int^\Lambda\!\!\frac{d^2 p}{(2\pi)^2}\frac{1}{p^2+(\omega_n-\alpha_*)^2+\sigma_*^2}\,,\\
\label{GNgap2}
i\frac{Q}{N}&=\lim_{\epsilon\rightarrow 0}\frac{2}{\b}\int^\Lambda\!\!\frac{d^2 p}{(2\pi)^2}\sum_{n=-\infty}^\infty\frac{e^{i\omega_n\epsilon}(\omega_n-\alpha_*)}{p^2+(\omega_n-\alpha_*)^2+\sigma_*^2}\,,
\end{align}
where  the fermionic Matsubara sums are over the discrete frequencies $\omega_n=(2n+1)\pi/\b$. As we explain in the Appendix, we have used the parameter $\epsilon$ to regulate the sum before performing the integral. The latter is also regulated using the UV cutoff $\Lambda$. 

At this moment is it important to recall the physics of the model in the absence of the chemical potential \cite{Rosenstein:1990nm,Moshe:2003xn}. Setting $\a_*=0$ into the first gap equation (\ref{GNgap1}) one finds 
\be
\label{GNgap0chem}
2\sigma_*\left[-\frac{{\cal M}}{4\pi}+\frac{\sigma_*}{4\pi}+\frac{1}{2\pi\b}\ln\left(1+e^{-\b\sigma_*}\right)\right]=0\,,
\ee
where
\be
\label{MGcrit}
\frac{{\cal M}}{4\pi}=\frac{1}{2G_*}-\frac{1}{2G}\,,\,\,\,\frac{1}{2G_*}=\int^\Lambda\!\!\!\frac{d^3p}{(2\pi)^3}\frac{1}{p^2}\,.
\ee
In writing (\ref{GNgap0chem}) we have taken the cutoff to infinity. The arbitrary mass scale ${\cal M}$ quantifies the distance of the bare coupling $G$ from the zero temperature critical coupling $G_*$. A non-zero solution for $\sigma_*$ of the gap equation (\ref{GNgap0chem}) would break parity by giving mass to the elementary fermions. This is only possible if ${\cal M}>0$, which requires $G>G_*$. But even then, the value of $\sigma_*$ goes to zero if the temperature reaches a critical value $T_c=1/\b_c = {\cal M}/2\ln 2$. At $G\leq G_*$ parity is restored since the only solution of (\ref{GNgap0chem}) is $\sigma_*=0$. Exactly at the critical point $G=G_*$ one can evaluate the (subracted) free energy density of the system as
\be
\label{feGN0}
f_f(\infty) -f_f(\b)\equiv \D f_f(\b)=N\frac{3}{2}\frac{\zeta(3)}{2\pi\beta^3}\,,
\ee
where $f(\b)=-\ln Z(\b)/(\b V_2)$ . The result (\ref{feGN0}) is identical with the corresponding thermal free energy density of $N$ free massless Dirac fermions in three dimensions reviewed in the Appendix.  

The physics of the system at real non zero chemical potential $\m$ has been studied in various works in past e.g. \cite{Hands:1992ck}. A real chemical potential lowers the critical temperature $T_c$ which eventually becomes zero at some critical value $\m_c$. Moving on to imaginary chemical potential, however, we discover some new features \cite{Christiansen:1999uv}. Firstly, the gap equation (\ref{GNgap1}) now becomes
\be
\label{GNgap11}
2\sigma_*\left[-\frac{{\cal M}}{4\pi}+\frac{\sigma_*}{4\pi}+\frac{1}{4\pi\b}\ln\left(1+2\cos(\b\alpha_*)e^{-\b\sigma_*}+e^{-2\b\sigma_*}\right)\right]=0\,.
\ee
At the critical point where ${\cal M}=0$, the gap equation (\ref{GNgap11}) has two  real non-zero solutions for $\s_*$ inside the window\footnote{We take $\b\a_*\in [-\pi,\pi]$.} $2\pi/3< |\b\a_*|< \pi$. These are the roots of 
\be
\label{eq}
x^2+(2\cos(\b\alpha_*)-1)x+1=0\,,\,\,\,x=e^{-\b\sigma_*}\,.
\ee
The above chemical potential window corresponds to a novel scaling regime of the fermionic system. In contrast to the case of zero or real chemical potential case, here when the system  is tuned to its zero temperature critical coupling $G_*$  it can break parity since $\s_*\neq 0$, while being at the same time  a thermal CFT. This is exactly analogous to the finite temperature behaviour of a three dimensional bosonic CFT as we will also review later \cite{Sachdev:1993pr,Petkou:1998wd}. 

The second gap equation (\ref{GNgap2}) is evaluated as\footnote{The following correct some misprints in \cite{Christiansen:1999uv}.} 
\be
\label{GNgap21}
\frac{Q}{N} =-\frac{i}{\pi\b^2}D(-e^{-\b\sigma_*-i\b\a_*})=\frac{i}{2\pi\b^3}\left[Cl_2(2\b\a_*)+Cl_2(2\phi-2\b\a_*)-Cl_2(2\phi)\right]\,.
\ee
 We notice here the remarkable appearance of the Bloch-Wigner function $D(z)$ \cite{Zagier1} in the result. This function is defined as
 \be
 \label{BW}
 D(z) = {\rm Im}[Li_2(z)]+\ln|z|{\rm Arg}(1-z)\,,
 \ee
 and can be expressed in terms of the Clausen function $Cl_2(z)$ (see p.8 of \cite{Kirillov:1994en})  with
\be
 \label{phidef}
 \phi=\arctan\left[\frac{e^{-\b\sigma_*}\sin(\b\a_*)}{1+e^{-\b\sigma_*}\cos(\b\a_*)}\right]\,.
 \ee
 It is well known e.g. \cite{Zagier1} that the Bloch-Wigner  function $D(z)$ gives the volume of an ideal tetrahedron in Euclidean hyperbolic space ${\cal H}_3$, when the four vertices of the former lie in $\partial{\cal H}_3$ at the points $ 0, 1,\infty,$ and $z$ ($z$ is a dimensionless cross ratio here). $D(z)$ is single-valued, real and analytic for all $z\in\mathbb{C}$, except at $z=0,1$ where it is non differentiable. We conclude that in this model there are no real solutions for $Q$, in accordance with our generic expectations mentioned in Section 2.
 
 The maximum value of $D(z)$ is obtained on the unit circle, for $z_{\pm}=(1\pm i\sqrt{3})/2$. These correspond to the end points of the interval discussed below (\ref{GNgap11}) $\b\a_*=\pm 2\pi/3$ where $\sigma_*=0$.  We obtain
 \be
 \label{Qmax}
 \frac{1}{N}Q_{max}\left(\pm\frac{2\pi}{3}\right)=\mp\frac{i}{\pi\b^2}Cl_2\left(\frac{\pi}{3}\right)\,.
 \ee
$Cl_2(\pi/3)=-Cl_2(-\pi/3)$ are the maximum (minimum) values of the Clausen function.

Recall now that for $\a_*=0$ and $\sigma_*=0$ we return to the zero chemical potential case, with $Q=0$ and the free energy density given by (\ref{feGN0}). However, at the middle point of the allowed interval $\b\a_*=\pi$, $\phi =0$ and as consequence we have $Q=0$.  At this point, the critical (${\cal M}=0$) gap equation becomes
\be
\label{GNgapbos}
2\sigma_*\left[\frac{\sigma_*}{4\pi}+\frac{1}{2\pi\b}\ln\left(1-e^{-\b\sigma_*}\right)\right]=0\,.
\ee
Interestingly, this can be written as
\be
\label{D1}
\frac{\s_*}{\p\b}D_1(e^{-\b\s_*})=0\,,
\ee
where $D_1(z)$ is the first in the series of the odd-indexed generalized Bloch-Wigner functions $D_m(z)$, $m=0,1,2,3...$ introduced by Zagier in \cite{Zagier2}. This has the real positive root 
\be
\label{gmean}
\sigma_*=\sigma_g\equiv\frac{2}{\b}\ln\left(\frac{1+\sqrt{5}}{2}\right)\,,
\ee
which can be interpreted as a parity breaking mass term for the fermions. Notice also that $\b\a_*=\pi$  results in the bosonization of the Matsubara frequencies in (\ref{GNgap1}) and (\ref{GNgap2}). At this value we find for the free energy density of the model 
\be
\label{GNfe}
\D f_f(\b)=-N\frac{8}{5}\frac{\zeta(3)}{2\pi \b^3}\,.
 \ee
 This is two times {\it minus} the large-$N$ free energy density of the bosonic $O(N)$ vector model at its non-trivial critical point \cite{Sachdev:1993pr,Petkou:1998wd}. The minus sign in (\ref{GNfe}) implies that the theory at $\b\a_*=\pi$ is non-unitary and it probably requires an analytic continuation to be matched with the usual $O(N)$ bosonic model.

\subsection{The CP$^{N-1}$ model at imaginary chemical potential}

The Euclidean action for the CP$^{N-1}$ model \cite{Arefeva:1980ms,DiVecchia} with an imaginary chemical potential is
\be
\label{CPaction}
S_{CPN}=\int_0^\b \!\!\!d\tau \!\!\int \!\!d^2x\left[|(\partial_\tau-i\a)\phi^a|^2 +|\partial_i\phi|^2 +i\sigma(\bar{\phi}^a\phi^a-\frac{N}{g})+iqa\right]\,,\,\,\,a=1,2,..,N\,,
\ee
where the auxiliary scalar field $\sigma$ enforces the constraint $|\phi|^2=N/g$ and $q$ is the constant eigenvalue density of the $U(1)$ charge density operator $\hat{q}=-ig\bar\phi^a\overset{\leftrightarrow}\partial_0\phi^a$. The model has a global $SU(N)$ symmetry, as well as a global $U(1)$ symmetry that can be trivially gauged by the introduction of a non-propagating abelian gauge field. Integrating out the scalar fields we obtain the canonical partition function as
\begin{align}
\label{CPPF1}
Z_b(\b,q)&=\int({\cal D}\alpha)({\cal D}\sigma)e^{-S_{b,eff}}\,,\\
\label{CPPF2}
S_{b,eff}&= iq\int_0^\b\!\!\!d\tau \!\!\int \!\!d^2x\,\alpha+iN\frac{1}{g}\int_0^\b\!\!\!d\tau\!\!\int \!\!d^2x\,\sigma-N\Tr\ln\left(-(\partial_0-i\a)^2-\partial^2+i\sigma\right)_\b\,.
\end{align}
Again, we  evaluate (\ref{CPPF1}) at constant saddle points $i\sigma_*\equiv m_*^2$ and $\a_*$. The latter are determined  by the gap equations
\begin{align}
\label{CPgap1}
\frac{1}{g}&=\frac{1}{\b}\sum_{n=-\infty}^\infty\int\frac{d^2 p}{(2\pi)^2}\frac{1}{p^2+(\omega_n-\alpha_*)^2+m_*^2}\,,\\
\label{CPgap2}
i\frac{q}{N}&=-\lim_{\epsilon\rightarrow 0}\frac{2}{\b}\int\frac{d^2 p}{(2\pi)^2}\sum_{-\infty}^\infty\frac{e^{i\omega_n\epsilon}(\omega_n-\alpha_*)}{p^2+(\omega_n-\alpha_*)^2+m_*^2}\,,
\end{align}
where the bosonic  frequencies are $\omega_n=2\pi n/\b$.

As before we recall briefly the phase structure of the model at zero chemical potential \cite{DiVecchia,Murthy:1989ps}. The gap equation (\ref{CPgap1}) reads in that case
\be
\label{CPgap0chem}
-\frac{{\cal M}}{4\pi}+\frac{m_*}{4\pi}+\frac{1}{2\pi\b}\ln\left(1-e^{-\b m_*}\right)=0\,,\,\,\,\,m_*>0\,,
\ee
where
\be
\label{MGcrit}
\frac{{\cal M}}{4\pi}=\frac{1}{g_*}-\frac{1}{g}\,,\,\,\,\frac{1}{g_*}=\int^\Lambda\!\!\!\frac{d^3p}{(2\pi)^3}\frac{1}{p^2}\,.
\ee
The gap equation (\ref{GNgap0chem}) diverges for $m_*=0$ which implies the absence of a finite temperature phase transition for the continuous $SU(N)$ global symmetry, in accordance with the Mermin-Wagner-Coleman theorem. At the critical coupling $g=g_*$ however, ${\cal M}=0$ and (\ref{GNgap0chem}) has a real solution which is given by the "golden mean" value shown in (\ref{gmean}). At this point, the free energy density is evaluated to be
\be
\label{feCP0}
 f_b(\infty)-f_b(\b)\equiv\D f_b(\b) =N\frac{8}{5}\frac{\zeta(3)}{2\pi \b^3}\,.
 \ee
To recapitulate, in the absence of a chemical potential the fermionic GN model has a parity breaking finite temperature transition but it does not exhibit a non-trivial scaling regime (except at some some physically obscure imaginary mass \cite{Petkou:2000xx}). The bosonic CP$^{N-1}$ model on the other hand, does not have a finite temperature transition but it exhibits a non trivial scaling regime. One should therefore have guessed by now that the presence of the imaginary chemical potential, or equivalently a Chern-Simons field expanded around a monopole background, catalyses an interpolation between these two theories. Lets see explicitly how this happens from the bosonic side this time.

The gap equation (\ref{CPgap1}) gives
\be
\label{CPgap11}
-\frac{{\cal M}}{4\pi}+\frac{m_*}{4\pi}+\frac{1}{4\pi\b}\ln\left(1-2\cos(\b\alpha_*)e^{-\b m_*}+e^{-2\b m_*}\right)]=0\,.
\ee
At the critical coupling where ${\cal M}=0$ the above has one real positive solution for $m_*$ in the window $-\pi/3< \b a_*<\pi/3$.  This is the root of
\be
\label{eq1}
x^2-(2\cos(\b\alpha_*) -1)x+1=0\,,\,\,\,x=e^{-\b m_*}\,.
\ee
Notice that this range is the mirror image wrt to the vertical axis of the allowed range for a non-zero fermionic mass discussed after (\ref{GNgap11}). For example, the middle point in the fermionic case  (for $\b\a_*=\pi$) is mapped into the middle point here which is at $\b\a_*=0$ (mod $2\pi$). 

The second gap equation (\ref{CPgap2}) is now evaluated to
\be
\label{CPgap21}
\frac{q}{N} =\frac{i}{\pi\b^2}D(e^{-\b\sigma_*-i\b\a_*})=-\frac{i}{2\pi\b^3}\left[Cl_2(2\b\a_*)+Cl_2(2w-2\b\a_*)-Cl_2(2w)\right]\,,
\ee
 where 
\be
 \label{wdef}
 w=\arctan\left[\frac{e^{-\b\sigma_*}\sin(\b\a_*)}{e^{-\b\sigma_*}\cos(\b\a_*)-1}\right]\,.
 \ee
As before the maximum value of $q$ is also reached at the end points of the interval above where $m_*=0$. We find 
\be
\label{qmax}
\frac{1}{N}q_{max}\left(\pm\frac{\pi}{3}\right)=\mp\frac{i}{\pi\b^2}Cl_2\left(\frac{\pi}{3}\right)\,.
 \ee

For $\a_*=0$ we return to the zero chemical potential case where $q=0$, the $SU(N)$ symmetry is unbroken, the mass of the scalars is given by (\ref{gmean}) and the free energy is (\ref{feCP0}). But now, for $\pi/3 \leq |\b\a_*|\leq \pi$ the only allowed solution for the gap equation (\ref{CPgap11}) is $m_*=0$. This might appear to imply the finite temperature breaking of the continuous $SU(N)$ symmetry and the violation of the Mermin-Wagner-Coleman theorem. However, we do not believe that this is the case. The underlying theory has now been effectively fermionized and the broken symmetry must be a discrete one. It would be important to clarify this point further, nevertheless. At the middle point of this window we have $\b\a_*=\pi$, $w=0$ and hence $q=0$. The theory has been now fermionized and the calculation of the free energy density yields
\be
\label{CPfe}
\D f_b(\b)=-N\frac{3}{2}\frac{\zeta(3)}{2\pi \b^3}\,.
\ee
This is exactly {\it minus} the fermionic free energy result (\ref{GNfe}). 

\subsection{The free energy map}
We have seen above that in the presence of an imaginary chemical potential the gap equations of the fermionic and bosonic models map into each other. As a consequence, there is a corresponding map of the physical properties between the two models. The map can be translated into a precise statement about the free energy densities of the models. 

The fermionic free energy density is given by
\begin{align}
\label{GNfefin}
\frac{1}{N}\D f_f(\b) &= -\int \frac{d^3 p}{(2\pi)^3}\ln p^2 +iQ\a_*+\frac{1}{\b}\sum_{n=-\infty}^\infty\int\frac{d^2\bp}{(2\pi)^2}\ln(\bp^2+(\omega_n-a_*)^2+\sigma_*^2)-\frac{\s_*^2}{2G}\nonumber \\
&=\int \frac{d^3 p}{(2\pi)^3}\left[\ln\left(\frac{p^2+\s_*^2}{p^2}\right)-\frac{\s_*^2}{p^2+\s_*^2}\right]+\frac{\a_*}{\pi\b^2}D(-z_*)\nonumber\\
&+\frac{\s_*^2}{2\pi\b}Re\left\{\ln(1+z_*)\right\}+\frac{1}{\pi\b}\int_{\s_*}^\infty x\,dx\,Re\left\{\ln\left(1+e^{-\b x-i\b\a_*}\right)\right\}\,,
\end{align}
where 
\be
\label{z*}
z_*=e^{-\b\s_*-i\b\a_*}\,,
\ee
and in the second line we have substituted  the gap equations. A few more details on the calculation are given in the Appendix. The second line of (\ref{GNfefin}) brings into the result the cutoff dependent critical coupling $1/2G_*$. This can be subtracted in order to obtain a finite result for the critical theory. Then, quite remarkably, the finite result of the second line line in (\ref{GNfefin}) combined with the third line gives {\it exactly} the generalized Bloch-Wigner-Zagier function $D_3(z)$ which is defined as \cite{Zagier2}
\be
\label{D3}
D_3(z)=Re\left\{Li_3(z)\right\}-\ln|z|Re\left\{Li_2(z)\right\}-\frac{1}{2}\ln^2|z|Re\left\{\ln(1-z)\right\}+\frac{\ln^3 |z|}{12}\,.
\ee
Our final result reads
\be
\label{GNfefin1}
\frac{1}{N}\D f_f(\b) = \frac{\a_*}{\pi\b^2}D(-z_*)-\frac{1}{\pi\b^3}D_3(-z_*)\,.
\ee
The analogous calculation for the bosonic theory yields
\be
\label{CPfefin1}
\frac{1}{N}\D f_b(\b) = -\frac{\a_*}{\pi\b^2}D(z_*)+\frac{1}{\pi\b^3}D_3(z_*)\,,
\ee
where we have set $m_*=\s_*$. We see that the free energies are mapped into each other when $\b\a_*\mapsto \b\a_*\pm\pi$. Precisely we obtain
\be
\label{femap}
\frac{1}{N}\D f_f(\b)\Biggl|_{\b\a_*\pm \pi} +\frac{1}{N}\D f_b(\b)\Biggl|_{\b\a_*}=\pm\frac{1}{\b^3}D(z_*)\,.
\ee

\section{Discussion}

Motivated by the recent revival of $3d$ bosonization we studied here the fermionic $U(N)$ Gross-Neveu and the bosonic CP$^{N-1}$ models at finite temperature and imaginary chemical potential for a $U(1)$ charge.  We started by noticing that if the charge density vanishes for non-trivial values of the chemical potential, the underlying system could effectively undergo a statistical transmutation. We have also pointed out that the canonical partition function of the systems are  intimately related to the thermal partition function of abelian Chern-Simons coupled to matter in a monopole background, when a suitable mean-field approximation is assumed. One may think of the latter as the regime where the $U(1)$ charge density essentially corresponds to the monopole charge. 

One of our main results is that phase structure of the Gross-Neveu and CP$^{N-1}$ models are altered in the presence of the imaginary chemical potential. A novel scaling phase opens up for the Gross-Neveu model while a pseudo-broken phase of a continuous symmetry appears in the CP$^{N-1}$ model. The latter situation may appear to violate the Mermin-Wagner-Coleman theorem, but we believe that this is not the case since the system has been fermionized and the broken symmetry must be discrete. Thus, the phase structures of the models appear to be  mapped into each other. A further result of our analysis is the fact that the charge density for both models is purely imaginary and is given by the Bloch-Wigner function. This is rather tantalizing. The Bloch-Wigner function gives the volume of ideal tetraherda in ${\cal H}_3$ and apart from its great importance for the classification of 3-manifolds, it also gives the imaginary part of a complex Chern-Simons action. This fits rather nicely with our association of the $U(1)$ charge density to the Chern-Simons level.  Morever, the second gap equation can be identified with the $D_1(z)$ function introduced by Zagier in \cite{Zagier2}. Finally, we have calculated the free energy densities of our two models and found, again rather remarkably, that they are given in terms of the generalized Bloch-Wigner function $D_3(z)$ \cite{Zagier2}. Our results allowed us to make a precise duality statement between the free energies of the models in (\ref{femap}). Again, the Bloch-Wigner function $D(z)$ plays an important role here as it gives the sum of the fermionic and bosonic free energy densities at imaginary chemical potentials that differ by $i\pi$. 

Our work raises numerous questions that could shed more light into the nature of $3d$ fermion-boson duality. Firstly, we note that our approach to the canonical formalism closely resembles the physics of a quantum mechanical system in a periodic potential. Indeed, the imaginary chemical potential may be though of as the periodic coordinate and the charge density as the quasi momentum. Such a quantum mechanical systems exhibit a band structure which can be studied by restricting the quasi momentum to the first Brillouin zone. One cannot go very far without a detailed model at hand, however there are certain topological properties like e.g. the Zak phase \cite{Zak} which only depend on the band symmetry. It is then tempting to associate the fermion-boson duality with some symmetry properties of the band.

Another interesting point is the surprising, to us, relevance of the hyperbolic geometry for the fermion-boson map. The bosonic and fermionic systems have a very different phase structure for real chemical potential when the charge density is real for both systems. In that case we are not aware of a manifestation of the boson-fermion map. Introducing the imaginary chemical potential the charge density becomes purely imaginary and apparently evaluates the volume of ideal tetrahedra in ${\cal H}_3$. Since the latter is related to the imaginary part of the complex $SL(2,\mathbb{C})$ Chern-Simons action, our result seem to suggest that the canonical partition function for fermions and bosons in three dimensions is related to the full partition function of the above complex group. It would be important to quantify such a possible relationship, as well as to understand the meaning of the charge density extrema. Finally, the appearance of the $D_1(z)$ function in the gap equation and the $D_3(z)$ function in the free energy density is another tantalising result. There is a large mathematical literature for symmetry properties of these functions (also related to Nielsen's generalized polylogarithms see e.g.\cite{Borwein}), which consequently should be inherited by the free energy densities of fermions and bosons in three dimensions. Finally, there are strong indications \cite{FPVtoappear} that analogous results, namely the appearance of higher $D_m(z)$ functions, are relevant in studies of higher dimensional fermionic and bosonic critical system such as the finite temperature extensions of the models discussed in \cite{Fei:2014yja}. 

\section*{Acknowledgements}
The work of A. C. P. is partially supported by the MPNS–COST Action MP1210 “The String Theory Universe”. We would like to thank Ofer Aharony and Pierre Vanhove for a critical reading of the manuscript and very helpful remarks. 
\appendix

\section{Notation and useful results}
Following \cite{ZinnJustin:2002ru} we use two-component Euclidean Dirac spinors and a Hermitian representation for the gamma matrices as $\gamma_\m=\s_\m$ where $\m=0,1,2$. $\sigma_\m$ are the usual Pauli matrices with the definition $\gamma_0=\sigma_0\equiv \sigma_3$.  Latin indices run as $i=1,2$.

A standard integral used in the text is
\be
\label{integral1}
\int^\Lambda\!\!\!\frac{d^3 p}{(2\pi)^3}\frac{1}{p^2+\sigma_*^2}=\frac{\Lambda}{2\pi^2}-\frac{\sigma_*}{4\pi}+O(\sigma_*/\Lambda)\,.
\ee
The Matsubara sums can be done with the help of the Poisson sum formula
\be
\label{Poisson}
\sum_{n=-\infty}^\infty f(n)=\sum_{k=-\infty}^\infty\int_{-\infty}^\infty dx \,e^{-i2\pi kx}f(k)\,.
\ee
As an application we consider the sum that appears in the fermionic charge density gap equation (\ref{GNgap2})
\be
\label{Qsum1}
\lim_{\e\rightarrow 0}\sum_{n=-\infty}^\infty\frac{e^{i\omega_n\epsilon}(\omega_n-\alpha_*)}{p^2+(\omega_n-\alpha_*)^2+\sigma_*^2}\,.
\ee
Without the introduction of the convergence factor $e^{i\omega_n\epsilon}$, $\e>0$ the sum would be undetermined \cite{SilvaNeto:1998dk}. Doing then the integral in the rhs of (\ref{Poisson}) term by term we first note that the $n=0$ term vanishes and we obtain
\be
\label{Qsum2}
\lim_{\e\rightarrow 0}\sum_{n=-\infty}^\infty\frac{e^{i\omega_n\epsilon}(\omega_n-\alpha_*)}{p^2+(\omega_n-\alpha_*)^2+\sigma_*^2} = i\frac{\b}{2}\left(\frac{1}{1+e^{\b\sqrt{\bp^2+\s_*}+i\b\a_*}}-\frac{1}{1+e^{\b\sqrt{\bp^2+\s_*}-i\b\a_*}}\right)\,.
\ee
Integrating (\ref{Qsum2}) over the spatial momenta $\bp$ and using the definition of the dilogarithm
\be
\label{dilog}
Li_2(z)=-\int_0^z \frac{dw}{w}\ln(1-w)\,,
\ee
we obtain (\ref{GNgap21}). The vanishing of the $n=0$ mode above should be contrasted with the sum that appears in the first gap equation (\ref{GNgap1}), which is evaluated to
\begin{align}
\label{gapexample}
\frac{1}{\b}\sum_{n=-\infty}^\infty\int\frac{d^2 p}{(2\pi)^2}\frac{1}{p^2+(\omega_n-\alpha_*)^2+\s_*^2}  &= \int\frac{d^3 p}{(2\pi)^3}\frac{1}{p^2+\s_*^2}\nonumber\\
&\hspace{-1cm}-\frac{1}{2}\int \frac{d^2\bp^2}{(2\pi)^2}\frac{1}{\sqrt{\bp^2+\s_*^2}}\left(\frac{1}{1+e^{\b\sqrt{\bp^2+\s_*}+i\b\a_*}}\frac{1}{1+e^{\b\sqrt{\bp^2+\s_*}-i\b\a_*}}\right)\,.
\end{align}
The divergent first term in the rhs of the above, which is needed for the coupling constant renormalization, comes from the $n=0$ term in the sum.

\section{Thermal partition function for free scalars and fermions in $d=3$}
We review in some detail here the free field theory results for the free energy density of scalars and fermions in $d=3$. Consider the $O(N)$ invariant action of free massive scalars $\phi^i(x)$, $a=1,2,..,N$ in $d=3$
\be
\label{scalaraction}
{\cal I}_b=\int d^3 x\left(\frac{1}{2}\partial_\m\phi^a\partial_\m\phi^a +\frac{1}{2}m_b^2\phi^a\phi^a\right)\,.
\ee
We put he theory in Euclidean $S_1\times\mathbb{R}^2$ where $S_1$ has radius $L=\beta=1/T$. Imposing periodic boundary conditions as
\be
\label{omegabos}
p_\m=(\omega_n,\bp)\,,\,\,\bar{p}=(p^2,p^2)\,,\,\,\omega_n=\frac{2\pi}{\b}n\,,\,\,\,n=0,\pm1,\pm2,..\,.
\ee
The thermal free energy  density $f_b(\b)$ is  defined as 
\be
\label{Zb}
Z_b=\int({\cal D}\phi^i)\exp^{-\frac{1}{2}\int_0^\b dx^3\int d^2x\left[\phi^i(-\partial^2)\phi^i+m_b^2\phi^i\phi^i\right]}\equiv e^{-\beta V_2f_b(\b)}\,,
\ee
with $V_2$ the volume of $\mathbb{R}^2$. The interesting quantity is the difference
\be
\label{fbpd}
f_b(\infty)-f_b(\b)\equiv \Delta f_b(\b)=\frac{N}{2}\int\frac{d^3 p}{(2\pi)^3}\ln p^2-\frac{N}{2L}\sum_{n=-\infty}^\infty\int\frac{d^2\bp}{(2\pi)^2}\ln(\bp^2+m_b^2+\omega^2_n)\,,
\ee
which is expected to be positive in a stable theory since $f_b(\b)=-{\cal P}_b(\b)$ is the bosonic pressure density at temperature $T=1/\b$. We find (we give the result for general $d$)
\begin{align}
\label{fbcalc}
\frac{1}{N}\Delta f_b(\b) &= -\frac{1}{2}\int\frac{d^3p}{(2\pi)^3}\ln\left(\frac{p^2+m_b^2}{p^2}\right)-\frac{1}{\b}\int\frac{d^{d-1}p}{(2\pi)^{d-1}}\ln\left(1-e^{-\b\sqrt{\vec{p}^2+m_b^2}}\right)
\nonumber \\
 &=-\frac{\pi S_d}{2d(2\pi)^d}\left[\frac{m_b^d}{\sin\frac{\pi d}{2}}-4d\frac{S_{d-1}}{S_d}\frac{1}{\b^d}\int_0^{e^{-\b m_b}}\left(\ln^2x-m_b^2\b^2\right)^{\frac{d-3}{2}}\ln x\ln(1-x)\frac{dx}{x}\right]
 \end{align}
 with $S_d=2\pi^{d/2}/\Gamma(d/2)$ the $d$-dimensional solid angle. 
 The integral above can be evaluated in closed form for $d$ odd, in terms of Nielsen's generalized polylogarithms \cite{Kolbig,Borwein} (and eventually as a finite sum of polylogarithms which turn into a Bloch-Wigner-Zagier function after the introduction of a gauge field or an imaginary chemical potential \cite{FPVtoappear}.) For $d=3$ and $m_b=0$ we have
 \be
 \label{fbm0}
 f_b(\infty)-f_b(L)=\frac{N}{2\pi L^3}\zeta(3)
 \ee
 This is {\it half} the value of the free energy density for the free CP$^{N-1}$ model. 

For $N$ Dirac fermions $\psi^i$, $\bar{\psi}^a$, $a=1,2,..,N$ in  three Euclidean we have
\be
\label{diracaction}
{\cal I}_f=-\int d^3x(\bar{\psi}^i\gamma^\m\partial_\m\psi^i +m_f\bar{\psi}^i\psi^i)
\ee

The corresponding thermal free energy calculation for fermions yields
\be
\label{ff}
\frac{1}{N\Tr\mathbb{I}}\Delta f_f(\b)=-\frac{1}{2}\int\frac{d^3 p}{(2\pi)^3}\ln p^2+\frac{1}{2\b}\sum_{n=-\infty}^\infty\int\frac{d^2\bp}{(2\pi)^2}\ln(\bp^2+m_b^2+\omega^2_n)
\ee
where the fermionic frequencies are $\omega_n=\frac{\pi}{\b^2}(2n+1)$. We obtain
\begin{align}
\label{ffcalc}
\frac{1}{N\Tr\mathbb{I}}\D f_f(\b)&= \frac{1}{2}\int\frac{d^3p}{(2\pi)^3}\ln\left(\frac{p^2+m_f^2}{p^2}\right)+\frac{1}{\b}\int\frac{d^{d-1}p}{(2\pi)^{d-1}}\ln\left(1+e^{-L\sqrt{\bp^2+m_f^2}}\right)
\nonumber \\
&=\frac{\pi S_d}{2d(2\pi)^d}\left[\frac{m_f^d}{\sin\frac{\pi d}{2}}-4d\frac{S_{d-1}}{S_d}\frac{1}{\b^d}\int_0^{e^{-\b m_f}}\left(\ln^2x-m_f^2\b^2\right)^{\frac{d-3}{2}}\ln x\ln(1+x)\frac{dx}{x}\right]
\end{align}
For $d=3$ and $m_f=0$ we obtain
\be
\label{ffm0}
\D f_(\b)=\frac{1}{2\pi L^3}N\Tr\mathbb{I}\frac{3}{4}\zeta(3)
\ee
Since $\Tr\mathbb{I}=2$ here, the fermionic result is 3/2 times the bosonic one.

\bibliographystyle{JHEP}
\bibliography{Refs}

\end{document}